\documentclass[12pt,fleqn]{iopart}
\usepackage{times}
\usepackage{iopams}
\usepackage{amsmath}

\newcommand{\vpsi}{{\boldsymbol{\psi}}}
\newcommand{\vsigma}{{\boldsymbol{\sigma}}}

\newcommand{\vtheta}{{\boldsymbol{\theta}}}

\newcommand{\vx}{{\mathbf{x}}}
\newcommand{\vxh}{\hat{\mathbf{x}}}
\newcommand{\vy}{{\mathbf{y}}}
\newcommand{\vyh}{\hat{\mathbf{y}}}
\newcommand{\vu}{{\mathbf{u}}}
\newcommand{\vv}{{\mathbf{v}}}

\newcommand{\vhh}{\hat{\mathbf h}}
\newcommand{\vh}{{\mathbf h}}

\newcommand{\tm}{{\mathbf m}}
\newcommand{\tmh}{\hat{\mathbf m}}
\newcommand{\tu}{{\mathbf{u}}}
\newcommand{\tv}{{\mathbf{v}}}
\newcommand{\tw}{{\mathbf{w}}}

\newcommand{\tkh}{\hat{\mathbf k}}
\newcommand{\tk}{{\mathbf k}}

\newcommand{\vtpsi}{{\boldsymbol{\psi}}}

\newcommand{\vtx}{{\mathbf{x}}}
\newcommand{\vtxh}{\hat{{\mathbf x}}}
\newcommand{\vty}{{\mathbf{y}}}
\newcommand{\vtyh}{\hat{{\mathbf y}}}

\newcommand{\hh}{\hat{h}}
\newcommand{\mh}{\hat{m}}
\newcommand{\kh}{\hat{k}}
\newcommand{\xh}{\hat{x}}

\newcommand{\yh}{\hat{y}}

\newcommand{\mxi}{{\boldsymbol{\xi}}}
\newcommand{\crit}{{\mathrm{c}}}

\newcommand{\rmD}{{\mathrm{D}}}

\newcommand{\mC}{{\mathbf{C}}}

\newcommand{\mG}{{\mathbf{G}}}

\newcommand{\mJ}{{\mathbf{J}}}

\newcommand{\mR}{{\mathbf{R}}}
\newcommand{\Rt}{{\tilde R}}
\newcommand{\mRt}{{\tilde \mathbf{R}}}
\newcommand{\mS}{{\mathbf{S}}}

\newcommand{\bqh}{\hat{{\mathbf q}}}
\newcommand{\bq}{{\mathbf q}}
\newcommand{\qh}{\hat{q}}
\newcommand{\bQh}{\hat{{\mathbf Q}}}
\newcommand{\bQ}{{\mathbf Q}}
\newcommand{\bC}{{\mathbf C}}
\newcommand{\bG}{{\mathbf G}}
\newcommand{\bM}{{\mathbf M}}
\newcommand{\bS}{{\mathbf S}}
\newcommand{\Qh}{\hat{Q}}
\newcommand{\bKh}{\hat{\mathbf K}}
\newcommand{\bK}{{\mathbf K}}
\newcommand{\Kh}{\hat{K}}
\newcommand{\negk}{\kern-0.8em}
\newcommand{\qt}{{\tilde{q}}}

\newcommand{\avg}[1]{\left\langle #1 \right\rangle}

\newcommand{\avgs}[1]{\left\langle #1 \right\rangle_{\mbox{\lower 1ex \hbox{\hskip -0.2 em\small *}}}\strut}
\newcommand{\davg}[1]{\left\langle\kern-0.4ex\left\langle #1 
		\right\rangle\kern-0.4ex\right\rangle}
\newcommand{\davgf}[1]{\langle\kern-0.4ex\langle #1 
		\rangle\kern-0.4ex\rangle}

\newcommand{\goesto}{\rightarrow}

\newcommand{\cD}[1]{\mathrm{d}#1\:}
\newcommand{\Order}[1]{{\mathcal{O}}\left(#1\right)}

\newcommand{\stick}{{s^\prime}}

\newcommand{\erf}{\operatorname{erf}}
\newcommand{\ooN}{{N^{-1}}}

\renewcommand{\mod}{\,\mathrm{mod}\,}

\newcommand{\mone}{{\mathbf 1}}

\newlength{\cprodlena}
\newlength{\cprodlenb}
\newcommand{\cprod}[2]{\ensuremath{%
	\settowidth{\cprodlena}{{$#1$}}%
	\settowidth{\cprodlenb}{{$#2$}}%
	\ifthenelse{\lengthtest{\cprodlenb > \cprodlena}}%
		{\setlength{\cprodlena}{\cprodlenb}}{}%
	\text{\hbox to \cprodlena {\hfil$#1$}} \otimes
	\text{\hbox to \cprodlena {$#2$\hfil}}
}}
\newcommand{\half}{\frac{1}{2}}

\newcommand{\bd}{\begin{displaymath}}
\newcommand{\ed}{\end{displaymath}}
\newcommand{\beq}{\begin{equation}}
\newcommand{\eeq}{\end{equation}}

\newcommand{\plus}{{\kern-0.01em+\kern-0.01em}}
\newcommand{\minus}{{\kern-0.01em-\kern-0.01em}}
\newcommand{\mal}{{\kern-0.01em\cdot\kern-0.01em}}
\newcommand{\postscriptfile}[1]{\relax}

\newcommand{\rmh}{{\rm h}}
\newcommand{\rmhh}{\hat{{\rm h}}}
\newcommand{\room}{\rule[-0.5cm]{0cm}{1.0cm}}

\renewcommand{\nonumber}{\relax}

\begin{document}

\jl{1} \title[Phase Diagram and Storage Capacity of $\ldots$]{Phase
Diagram and Storage Capacity \\ of Sequence Processing Neural
Networks}

\author{A D\"uring\dag, A C C Coolen\ddag, and D Sherrington\dag}
\address{\dag\ Department of Physics, University of
Oxford, 1 Keble Rd., Oxford OX1 3NP, UK}
\address{\ddag\ Department of Mathematics, King's College,
University of London, Strand, London WC2R 2LS, UK}
\date{\today}

\begin{abstract}
We solve the dynamics of Hopfield-type neural networks 
which store sequences of patterns, close to
saturation. The asymmetry of the interaction matrix in such models leads to
violation of detailed balance, 
ruling out an equilibrium statistical mechanical analysis.
Using generating functional methods we
derive exact closed equations
for dynamical order parameters, viz. the sequence overlap and
correlation- and response functions, in the thermodynamic limit.
We calculate the time translation
invariant solutions of these equations, describing stationary
limit-cycles, which leads to a phase diagram. The
effective retarded self-interaction usually appearing in symmetric models 
is here found to vanish, 
which causes a significantly enlarged storage capacity of
$\alpha_\crit\!\sim 0.269$, compared to $\alpha_\crit\sim 0.139$ for
Hopfield networks storing static patterns. Our 
results are tested against extensive
computer simulations and excellent agreement is found.
\end{abstract}
\submitted
\maketitle

\setlength{\mathindent}{\parindent}

\section{Introduction}

The equilibrium  properties of the Hopfield model \cite{hopfield82}, a
globally coupled neural network, with the typically Hebbian prescription for
the interaction strengths
\begin{equation}
J_{ij}=\frac{1}{N}\sum_{\mu=1}^p\xi^{\mu}_i\xi^\mu_j,
\label{jij_standard}
\end{equation}
(in which the $\xi_i^\mu$ represent components of patterns to be
stored) have been successfully described in the regime
close to saturation, where the number $p$ of patterns stored scales as
$p=\alpha N$, using replica methods \cite{amit85,amit87}. As an
alternative approach, a path integral formalism developed in 
\cite{dedominicis78}, was applied to the dynamics 
of the same system, and both approaches have indeed been
shown to lead to identical phase diagrams \cite{rieger88}.  Many
modifications of the standard Hopfield model have been proposed,
including models where the network does not statically recall
individual patterns, but reproduces a {\em sequence\/} of stored patterns
\cite{hopfield82,sompolinsky86,amari88a,bauer90,kuehn91,coolen92}. The
simplest way to induce this cyclical behaviour is by an
asymmetric modification of the interaction matrix
(\ref{jij_standard}), in combination with a parallel execution of the
neural dynamics. Numerical simulations show that in such
(non-symmetric) models the number of patterns that can be stored 
successfully is significantly larger than that of models 
storing static patterns, 
with a storage capacity of $\alpha_\crit\approx
0.27$ \cite{hopfield82,amari88a}, compared to $\alpha_\crit\approx 0.14$ for
the standard (symmetric) Hopfield model.

In this paper we study such a model, where    
a single sequence of extensive length is stored in a fully 
(but non-symmetrically) connected 
Ising spin neural network with parallel stochastic dynamics. 
The asymmetry of the interaction matrix and the resulting violation of
detailed balance and associated fluctuation-dissipation theorems rule 
out equilibrium statistical mechanical methods
of analysis, including conventional replica theory. Some time ago 
an approximate dynamical solution for this model was proposed  
\cite{amari88a}, which provided results roughly in line 
with the numerical evidence available at the time. To our knowledge, 
an exact solution, however, has so far not yet been found.   

In our present study we use the path integral methods of
\cite{dedominicis78,rieger88,sommers87} to solve the dynamics close to
saturation exactly in the thermodynamic limit for our fully connected
sequence processing network, 
without having to resort to approximations. 
In the standard (symmetric) Hopfield network two effects
limit the storage capacity: a Gaussian noise in the equivalent
effective single spin problem, which is non-local in time, and a
retarded self-interaction. The magnitude of both depends
on the load factor $\alpha$. Our theory shows 
that for the present model the
retarded self-interaction vanishes, similar to the situation
in the non-symmetric SK model \cite{crisanti87,crisanti88,rieger89},
which explains the extended storage capacity. Numerical simulations
for large system sizes (up to 50,000 spins) are in excellent agreement
with our analytical results, both with respect to the maximum storage
capacity $\alpha_{\text{c}}\approx 0.269$ (at zero noise level) and
with respect to the full phase diagram in the $\alpha-T$ plane. 

\section{Definitions}

We study a system consisting of $N$ Ising-type neurons 
$\sigma_i=\pm 1$ which evolve in time according to a stochastic
alignment to local fields.  The neurons change their states 
simultaneously, with probabilities
\begin{equation}
{\rm Prob}[\sigma_i(t\!+\!1)\!=\!-\sigma_i(t)]~=~
\frac{1}{2}\left[1-
\tanh\!\left(\!
\beta\sigma_i(t)\left[\sum_{j=1}^NJ_{ij}\sigma_j(t)\!+\!\theta_i(t)\right]\!\right)\!\right],
\label{flip_prob}
\end{equation}
\noindent where the entries of the interaction matrix $\mJ$ are given by
\begin{equation}
J_{ij}=\frac{1}{N}\sum_{\mu=1}^p\xi^{\mu+1}_i\xi^\mu_j
\label{jij}
\end{equation}
(the pattern labels $\mu$ are understood to be taken modulo
$p$).  The non-negative parameter $\beta=T^{-1}$ controls the amount
of noise in the dynamics, with $T=0$ corresponding to deterministic
evolution and with $T=\infty$ corresponding to purely random
evolution.  The variables $\theta_i(t)$ represent external fields. 
The $p$ vectors $\bxi^\mu=(\xi^\mu_1,\ldots
,\xi_N^\mu)\in\{-1, 1\}^{N}$ are randomly and independently drawn 
 patterns. Our
interest is in the saturation regime $p=\alpha N$. 
For discussions of the relation of
such models to biological or artificial neural networks see e.g. 
\cite{amitbook,hertz91,peretto92,sherrington93}. 
The matrix $\mJ$ will generally be
non-symmetric, so that (\ref{flip_prob}) will not obey detailed
balance. Hence we can not use conventional equilibrium statistical
mechanics to analyze the stationary behaviour: we will have to solve
the dynamics. For the subsequent analysis, it will
turn out to be useful to represent our expression for $\mJ$ in matrix
notation as
\begin{equation}
\mJ=\frac{1}{N}\left(\mxi^T\mS\mxi\right),\qquad
S_{\mu\nu}=\delta_{\mu, (\nu +1) \mod p}
\label{mJ_def}
\end{equation}
\noindent Here the $p\times N$ matrix $\mxi$ is defined as $\mxi_{\mu
i}=\xi^\mu_i$. 
When $\mS$ is replaced by the unity matrix $\mone$, the definition (\ref{mJ_def})
reverts to that of the standard Hopfield model. 

To analyze the dynamics of the system we follow \cite{dedominicis78}
and define a generating (or characteristic) functional $Z[\vtpsi]$:
\begin{equation}
Z[\vtpsi]=\!\sum_{\vsigma(0)\ldots\vsigma(t)}
        p[\vsigma(0),\ldots,\vsigma(t)]
       ~\rme^{~ -\rmi\sum_{s<t}\vsigma(s)\mal\vpsi(s)},
\label{Z_def}
\end{equation}
\noindent in which $\vsigma(s)=(\sigma_1(s),\ldots,\sigma_N(s))$
denotes the microscopic system state at time $s$, and with the usual
notation $\vx\mal\vy=\sum_{i}x_iy_i$. In the familiar way one can
obtain 
from $Z[\vtpsi]$ all averages of interest by differentiation, e.g.  
\begin{align}
m_i(s)&=
\avg{\sigma_i(s)}=\rmi\lim_{\vtpsi\goesto 0} \frac{\partial
Z[\vtpsi]}{\partial \psi_i(s)}
\nonumber \\
G_{ij}(s,\stick)&=\frac{\partial}{\partial\theta_j(\stick)}\avg{\sigma_i(s)}=\rmi\lim_{\vtpsi\goesto 0}
     \frac{\partial^2 Z[\vtpsi]}{\partial \psi_i(s)\partial
\theta_j(\stick)}
\nonumber \\
C_{ij}(s,\stick)&=\avg{\sigma_i(s)\sigma_j(\stick)}=-\lim_{\vtpsi\goesto 0}
        \frac{\partial^2 Z[\vtpsi]}{\partial \psi_i(s)\partial
\psi_j(\stick)}.
\nonumber
\end{align}
The dynamics (\ref{flip_prob}) is a Markov chain, so 
the path probabilities $p[\vsigma(0),\ldots,\vsigma(t)]$ are simply 
given by products of the individual transition probabilities 
$W[\vsigma^\prime|\vsigma]$ of the
chain:  
$p[\vsigma(0),\ldots,\vsigma(t)]=p[\vsigma(0)]\prod_{s=0}^{t-1}W[\vsigma(s+1)|\vsigma(s)]$. For the
dynamics (\ref{flip_prob}) these transition probabilities 
are given by 
\bd
 W[\vsigma(s+1)|\vsigma(s)]=
\prod_{i}~
\frac{1}{2}\left[1+\sigma_i(s\plus 1)
\tanh(
\beta[\sum_{j}J_{ij}\sigma_j(t)\!+\!\theta_i(t)])\right]
\ed
\vspace*{-9mm}

\bd
~~~~~~~~~~~~~~~~~~~~=
\prod_{i}
 \rme^{~\beta\sigma_i(s+1)[\sum_{j}J_{ij}\sigma_j(s)
                +\theta_i(s)]-\ln
        2\cosh(\beta[\sum_{j}J_{ij}\sigma_j(s)+\theta_i(s)])}
\ed
To formally remove the coupling terms
$\sigma_i(s+1)\sigma_j(s)$ we introduce the auxiliary variables
$\vh(s)=(h_1(s),\ldots,h_N(s))$, representing the local fields at
each spin site at given times, by insertion of
\bd
1=\int{\rmd\vh(s)}\prod_{i}\delta[h_i(s)-\sum_{j}J_{ij}\sigma_j(s)
                -\theta_i(s)]. 
\ed
After writing the above $\delta$-distributions in integral form, which
generates conjugate field variables 
$\vhh(s)=(\hat{h}_1(s),\ldots,\hat{h}_N(s))$, and upon introducting
the more convenient notation 
$\{\rmd\vh\rmd\vhh\}=\prod_{i}\prod_{s<t}[dh_i(s)d\hat{h}_i(s)/2\pi]$,  
we can express (\ref{Z_def}) as
\bd
Z[\vtpsi]
=\!\! \!\sum_{\vsigma(0)\ldots\vsigma(t)}\!\! 
p(\vsigma(0))\!\int\! \{\rmd\vh \rmd\vhh\}
\prod_{s<t}
\rme^{~\beta\vsigma(s+1)\mal\vh(s)
        -\sum_{i}\ln 2\cosh[\beta
h_i(s)]+\rmi\vhh(s)\mal[\vh(s)-\vtheta(s)]
-\rmi\vpsi(s)\mal\vsigma(s)}
\ed
\vspace*{-12mm}

\begin{equation}
~~~~~~~~~~~~~~~~~~~~~~~~~~~~~~~~~~~~~~~~~~~~~~~~~~
~~~~~~~~~~~~~~~~~~~~~~~~~~~~~~~\times~
\rme^{~-\rmi\ooN\vhh(s)\mal(\mxi^T\mS\mxi)\vsigma(s)}
\label{last_before_condensed}
\end{equation}
This expression describes the system dynamics
(\ref{flip_prob},\ref{jij}) in general form.  To obtain
quantitative information about particular regimes of operation, we
have to make specific ans\"atze.
Our ansatz will be one describing (possibly noisy) recall of
the stored sequence of patterns.  At each time step exactly one stored
pattern is assumed to be `condensed', i.e. the overlap between that
pattern (which without loss of generality
can be labelled with the time index) and the system state is of order
${\cal O}(1)$, whereas all other overlaps are of order ${\cal
O}(N^{-\frac{1}{2}})$. 
The cumulative impact of the overlaps of the
non-condensed 
patterns will introduce 
an additional noise component into the system dynamics; the
non-condensed patterns play the role of `quenched disorder'. 
For $N\to\infty$ the mean-field physics of the problem should be
self-averaging with respect to the realisation of the disorder, so
we are allowed to average the generating
functional (\ref{last_before_condensed}) over the non-condensed
patterns (such averages will be
denoted as $\overline{f[\{\mxi\}]}$). 
Since each pattern with $\mu\leq t$ will at some
stage be condensed, in contrast to those patterns with $\mu>t$, we can
simplify our calculation by 
averaging only over the latter. The resulting expressions will, for
$N\to\infty$, turn out not 
to depend on the remaining patterns with $\mu\leq t$. 

As in most dynamic mean-field calculations of disordered systems based 
on evaluating disorder-averaged 
generating functionals, we will consider the time $t$ to be
fixed, whereas we will take the limit $N\to\infty$. This restricts the 
predicting power of the theory to those processes that take
place on finite time-scales. In the present calculation we will
find that a time-translation invariant state (representing motion on a
stationary limit-cycle) is indeed approached on finite time-scales, so this 
restriction is not a problem. 

\section{Dynamic Mean Field Theory}

In (\ref{last_before_condensed}) only the term 
$\vhh(s)\mal(\mxi^T\mS\mxi)\vsigma(s)$ contains 
both condensed and non-condensed patterns. 
We isolate the non-condensed ones by introducing the
variables $\vx$ and $\vy$:
\bd
1=
\int\!\cD{\vtx}
\prod_{s<t}\prod_{\mu\neq s}\delta\left[x_\mu(s)\minus 
        \frac{1}{\sqrt{N}}\sum_{i}
                \xi^{\mu+1}_i\hh_i(s)\right]
\ed
\vspace*{-10mm}

\bd
1=\int\!\cD{\vty}
\prod_{s<t}\prod_{\mu\neq s}
\delta\left[y_\mu(s)\minus 
        \frac{1}{\sqrt{N}}\sum_{i}
                \xi^{\mu}_i\sigma_i(s)\right]
\ed
Upon writing the above $\delta$-distributions in integral form (which 
generates the additional integration variables  $\vxh$ and $\vyh$),  
we then
arrive at the following expression for the disorder-averaged
generating functional: 
\bd
\overline{Z}[\vtpsi]=\!\! \!\sum_{\vsigma(0)\ldots\vsigma(t)}\!\! 
p(\vsigma(0))\!\int\! \{\rmd\vh \rmd\vhh\}
~ 
\rme^{~\sum_{s<t}\left[\beta\vsigma(s+1)\mal\vh(s)
        -\sum_{i}\ln 2\cosh[\beta
h_i(s)]+\rmi\vhh(s)\mal[\vh(s)-\vtheta(s)]
-\rmi\vpsi(s)\mal\vsigma(s)\right]}
\ed
\vspace*{-10mm}

\bd
\times~
\rme^{~-i\ooN\sum_{s<t}[\vhh(s)\mal\mxi^{s+1}][\vsigma(s)\mal\mxi^{s}]}
\int\!\frac{\cD{\vtx}\cD{\vtxh}\cD{\vty}\cD{\vtyh}
}{(2\pi)^{2(p-1)t}}~\rme^{~\rmi\sum_{s<t}\sum_{\mu\neq s}\left[
\xh_\mu(s)x_\mu(s)+\yh_\mu(s)y_\mu(s)-x_\mu(s) y_\mu(s)\right]}
\ed
\vspace*{-8mm}

\begin{equation}
~~~~~~~~~~~~~~~~~~~~~~~~~~~~~~~~~~~~~~~~~~~~~~~~~~
\times~
\overline{\left[\rme^{~-\rmi N^{-\frac{1}{2}}\sum_{s<t}\sum_{\mu\neq s}
\left[\xh_\mu(s)\vhh(s)\mal\mxi^{\mu+1}+
\yh_\mu(s)\vsigma(s)\mal\mxi^{\mu}\right]}\right]}
\label{last_before_avg}
\end{equation}
We can now carry out the disorder average in the last term, which is significantly 
simplified if in the exponent we use  
$\sum_{\mu\neq s}[\sum_i \ldots] =\sum_{\mu>t}[\sum_i\ldots] +{\cal O}(N)$. It gives 
\bd
\overline{\left[{}^{}_{}\ldots{}^{}_{}\right]}
=\rme^{~{\cal O}(N^{\frac{1}{2}})}~ \overline{\left[e^{-\rmi
N^{-\frac{1}{2}}\sum_{s<t}\sum_{\mu>t}\sum_i \xi_i^{\mu}
\left[\xh_{\mu-1}(s)\hh_i(s)+
\yh_\mu(s)\sigma_i(s)\right]}\right]}
\ed
\vspace*{-10mm}

\bd
~~~~
=\rme^{~{\cal O}(N^{\frac{1}{2}})}~ \prod_{\mu>t}\prod_i~\cos\left[
N^{-\frac{1}{2}}\sum_{s<t}[\xh_{\mu-1}(s)\hh_i(s)+
\yh_\mu(s)\sigma_i(s)]\right]
\ed
\vspace*{-9mm}

\begin{equation}
~~~~
=\rme^{~{\cal O}(N^{\frac{1}{2}})}~ \prod_{\mu>t}
\rme^{~-\frac{1}{2N}\sum_{s,s^{\prime}<t}\sum_i
[\xh_{\mu-1}(s)\hh_i(s)+\yh_\mu(s)\sigma_i(s)]
[\xh_{\mu-1}(s^\prime)\hh_i(s^\prime)+\yh_\mu(s^\prime)\sigma_i(s^\prime)]}
\label{disorder_avg}
\end{equation}
Since the leading order $N$ in the exponent of (\ref{disorder_avg})
does not involve components of $\{\vx,\vxh,\vy,\vyh\}$ with pattern
index $\mu\leq t$, the latter can be integrated out in expression
(\ref{last_before_avg}).     
We now isolate the various relevant macroscopic 
observables occurring in (\ref{disorder_avg})  by inserting  
integrals over appropriate $\delta$-functions:
\bd
1=\int\!\frac{\cD{\tm}\cD{\tmh}}{(2\pi/N)^{t}}~
\rme^{~\rmi N\sum_{s<t}\mh(s)\left[m(s)-
\frac{1}{N}\sum_{i}\xi_i^{s}\sigma_i(s)\right]}
\ed
\vspace*{-10mm}

\bd
1=\int\!\frac{\cD{\tk}\cD{\tkh}}{(2\pi/N)^{t}}~
\rme^{~\rmi N\sum_{s<t}\kh(s)\left[k(s)-
\frac{1}{N}\sum_{i}\xi_i^{s+1}\hh_i(s)\right]}
\ed
\vspace*{-10mm}

\bd
1=\int\!\frac{\cD{\bq}\cD{\bqh}}{(2\pi/N)^{t^2}}~
\rme^{~\rmi N\sum_{s,\stick<t}\qh(s,\stick)\left[q(s,\stick)-
                \frac{1}{N}\sum_i\sigma_i(s)
                        \sigma_i(\stick)\right]}
\ed
\vspace*{-10mm}

\bd
1=\int\!\frac{\cD{\bQ}\cD{\bQh}}{(2\pi/N)^{t^2}}~
\rme^{~\rmi N\sum_{s,\stick<t}\Qh(s,\stick)\left[Q(s,\stick)-
                \frac{1}{N}\sum_i\hh_i(s)
                        \hh_i(\stick)\right]}
\ed
\vspace*{-10mm}

\bd
1=\int\frac{\cD{\bK}\cD{\bKh}}{(2\pi/N)^{t^2}}~
\rme^{~\rmi N\sum_{s,\stick<t}\Kh(s,\stick)\left[K(s,\stick)-
                \frac{1}{N}\sum_i\sigma_i(s)
                        \hh_i(\stick)\right]}
\ed
Combination of (\ref{disorder_avg}) with (\ref{last_before_avg})
will then give us an expression for $\overline{Z}[\vtpsi]$ which will 
factorise over sites if we choose a factorised initial distribution 
$p(\vsigma(0))=\prod_i p_i(\sigma_i(0))$, 
resulting in an integral which for $N\to\infty$ will be dominated by 
saddle-points: 
\begin{equation}
\overline{Z}[\vtpsi]=\int\!\cD{\tm}\cD{\tmh}\cD{\tk}\cD{\tkh}\cD{\bq}\cD{\bqh}
        \cD{\bQ}\cD{\bQh}\cD{\bK}\cD{\bKh}~\rme^{~N\left\{
        \Psi[\ldots]+
        \Phi[\ldots]+\Omega[\ldots]\right\}+
{\cal O}(N^{\frac{1}{2}})}
\label{decomposed}
\end{equation}
in which the functions $\Psi[\ldots]$, $\Phi[\ldots]$ and
$\Omega[\ldots]$ are given by:
\bd
\Psi[\tm,\tk,\tmh,\tkh,\bq,\bQ,\bK,\bqh,\bQh,\bKh]=
\rmi\sum_{s<t}\left[\mh(s)m(s)+\kh(s)k(s)-m(s)k(s)\right]
\ed
\vspace*{-11mm}

\begin{equation}
~~~~~~~~~~~~~~~~~~~
+\rmi\sum_{s, \stick<t}\left[
                        \qh(s,\stick)q(s,\stick)+
                        \Qh(s,\stick)Q(s,\stick)+
                        \Kh(s,\stick)K(s,\stick)
                \right]
\label{define_psi}
\end{equation}
\bd
\Phi[\tm,\tk,\bqh,\bQh,\bKh]=
\frac{1}{N}\sum_{i}\ln\left\{
\sum_{\sigma(0)\ldots\sigma(t)}\!p_i(\sigma(0))
\int\!\{\rmd\rmh \rmd\rmhh\}~\rme^{~\sum_{s<t}\left[
                \beta\sigma(s+1)h(s)
                -\ln 2\cosh[\beta h(s)]\right]}
\right.
\ed
\vspace*{-11mm}

\bd
\left.
~~~~~~~~~~~~~~~~~~~~~~~~~~~~~~~~~~~~~~
\times~
\rme^{~
-\rmi\sum_{s,\stick<t}\left[
                \qh(s,\stick)\sigma(s)\sigma(\stick)+
                \Qh(s,\stick)\hh(s)\hh(\stick)+
                \Kh(s,\stick)\sigma(s)\hh(\stick)\right]}
\right.
\ed
\vspace*{-12mm}

\begin{equation}
\left.
~~~~~~~~~~~~~~~~~~~~~~~~~~~~~~~~~~~~~~
\times~
\rme^{~\rmi \sum_{s<t}
\hh(s)[h(s)-\theta_i(s)-\kh(s)\xi_i^{s+1}]
-\rmi \sum_{s<t}\sigma(s)[\mh(s)\xi_i^s+\psi_i(s)]}
\room
\right\}
\label{define_phi}
\end{equation}
\bd
\Omega[\bq,\bQ,\bK]=
\frac{1}{N}\ln\int\!
\frac{\cD{\vtx}\cD{\vtxh}\cD{\vty}\cD{\vtyh}}{(2\pi)^{2(p-t)t}}~
\rme^{~\rmi\sum_{\mu>t}\sum_{s<t}\left[\xh_\mu(s)x_\mu(s)+\yh_\mu(s)y_\mu(s)-x_\mu(s)y_\mu(s)\right]}
\ed
\vspace*{-11mm}

\bd
~~~~~~\times~\rme^{~-\frac{1}{2}\sum_{\mu>t}\sum_{s,s^\prime<t}\left[
\xh_{\mu}(s)Q(s,s^\prime)\xh_{\mu}(\stick)
+\xh_{\mu\minus 1}(s)K(s^\prime,s)\yh_\mu(\stick)
+\yh_{\mu}(s)K(s,s^\prime)\xh_{\mu-1}(\stick)
+\yh_\mu(s)q(s,s^\prime)\yh_\mu(\stick)\right]}
\ed
\vspace*{-11mm}

\bd
=
\frac{1}{N}\ln\int\!
\frac{\cD{\tu}\cD{\tv}}{(2\pi)^{(p-t)t}}~
\rme^{~\rmi\sum_{\mu>t}\sum_{s<t}u_{\mu+1}(s)v_\mu(s)}
\ed
\vspace*{-11mm}

\begin{equation}
~~~~~~\times~\rme^{~-\frac{1}{2}\sum_{\mu>t}\sum_{s,s^\prime<t}\left[
u_{\mu}(s)Q(s,s^\prime)u_{\mu}(\stick)
+u_{\mu}(s)K(s^\prime,s)v_\mu(\stick)
+v_\mu(s)K(s,s^\prime)x_{\mu}(\stick)
+v_\mu(s)q(s,s^\prime)v_\mu(\stick)\right]}
\label{define_omega}
\end{equation}
with the short-hand
$\{\rmd\rmh\rmd\rmhh\}=\prod_{s<t}[dh(s)d\hat{h}(s)/2\pi]$.   
The final expression (\ref{define_omega}) for $\Phi$ was
obtained by integrating 
out the variables $(\vx,\vy)$, followed by a simple pattern index
shift transformation. 
\clearpage

One can deduce the physical meaning of the various dynamic order
parameters introduced along the way in the usual manner by (repeated) 
derivation of the definition (\ref{Z_def}) with respect
to the external fields $\theta_i(s)$ and $\psi_i(s)$, in combination
with usage of the normalisation identity $Z[{\bf 0}]=1$. 
Evaluation of a 
function $f[\ldots]$ at the dominating (physical) saddle-point of the  
extensive exponent in  (\ref{decomposed}) 
will be indicated by $f|_{\rm saddle}$. 
The external fields occur in the function $\Phi$ only (not in $\Psi$
or $\Omega$). 
The resulting identities can be summarised in a compact form upon
introduction of an effective single-site measure $\avg{\ldots}_{i}$,
defined as
\bd
\avg{f[\{\sigma,h,\hh\}]}_i= 
\frac{
\sum_{\sigma(0)\ldots\sigma(t)} \int\!\{\rmd\rmh \rmd\rmhh\}~W_i[\{\sigma,h,\hh\}]~f[\{\sigma,h,\hh\}]}
{\sum_{\sigma(0)\ldots\sigma(t)} 
\int\!\{\rmd\rmh \rmd\rmhh\}~W_i[\{\sigma,h,\hh\}]}
\ed
with 
\bd
W_i[\{\sigma,h,\hh\}]=p_i(\sigma(0))
\left.\left[
\rme^{~\sum_{s<t}\left[
                \beta\sigma(s+1)h(s)
                -\ln 2\cosh[\beta h(s)]
+\rmi\hh(s)[h(s)-\theta_i(s)-\kh(s)\xi_i^{s+1}]
-\rmi\sigma(s)\mh(s)\xi_i^s
\right]}
\right.\right.
\ed
\vspace*{-14mm}

\begin{equation}
~~~~~~~~~~~~~~~~~~~~~~~~~~~~~~
\left.\left.
\times~
\rme^{~-\rmi\sum_{s,\stick<t}\left[
                \qh(s,\stick)\sigma(s)\sigma(\stick)+
                \Qh(s,\stick)\hh(s)\hh(\stick)+
                \Kh(s,\stick)\sigma(s)\hh(\stick)\right]}
\right]\right|_{\rm saddle}
\label{single_site_measure}
\end{equation}
In particular we now find, in leading order in $N$:
\begin{equation}
\overline{\avg{\sigma_i(s)}}=\rmi\lim_{\vpsi\to
0}\left.\frac{\partial (N\Phi)}{\partial\psi_i(s)}\right|_{\rm saddle}
=\avg{\sigma(s)}_i
\label{first_link}
\end{equation} 
\vspace*{-10mm}

\begin{equation}
0=\frac{\partial\overline{Z}[{\rm 0}]}
{\partial\theta_i(s)}=\lim_{\vpsi\to
0}\left.\frac{\partial (N\Phi)}{\partial\theta_i(s)}\right|_{\rm
saddle}
=-\rmi \avg{\hh(s)}_i
\label{second_link}
\end{equation} 

\bd
\overline{\avg{\sigma_i(s)\sigma_j(s)}}=
-\lim_{\vpsi\to 0}
\left.\frac{\partial^2 (N\Phi)}{\partial\psi_i(s)\partial\psi_j(s^\prime)}\right|_{\rm saddle}
-\lim_{\vpsi\to 0}
\left.\left[\frac{\partial (N\Phi)}{\partial\psi_i(s)}
\frac{\partial (N\Phi)}{\partial\psi_j(s^\prime)}\right]\right|_{\rm saddle}
\ed
\vspace*{-10mm}

\begin{equation}
~~~~~~~~~~~~~~~~~~~~
=\delta_{ij}\avg{\sigma(s)\sigma(s^\prime)}_i+[1\minus\delta_{ij}]\avg{\sigma(s)}_i\avg{\sigma(s^\prime)}_i
\label{third_link}
\end{equation} 
\vspace*{-10mm}

\bd
\frac{\partial\overline{\avg{\sigma_i(s)}}}
{\partial\theta_j(s^\prime)}=
\rmi\lim_{\vpsi\to 0}
\left.\frac{\partial^2 (N\Phi)}{\partial\psi_i(s)\partial\theta_j(s^\prime)}\right|_{\rm saddle}
+\rmi\lim_{\vpsi\to 0}
\left.\left[\frac{\partial (N\Phi)}{\partial\psi_i(s)}
\frac{\partial (N\Phi)}{\partial\theta_j(s^\prime)}\right]\right|_{\rm
saddle}
\ed
\vspace*{-10mm}

\begin{equation}
~~~~~~~~~~~~~~~~~~~~
=-\rmi\delta_{ij}\avg{\sigma(s)\hh(s^\prime)}_i
\label{fourth_link}
\end{equation} 
\vspace*{-10mm}

\bd
0=\frac{\partial^2\overline{Z}[{\rm 0}]}
{\partial\theta_i(s)\partial\theta_j(s^\prime)}
=
\lim_{\vpsi\to 0}
\left.\frac{\partial^2 (N\Phi)}{\partial\theta_i(s)\partial\theta_j(s^\prime)}\right|_{\rm saddle}
+\lim_{\vpsi\to 0}
\left.\left[\frac{\partial (N\Phi)}{\partial\theta_i(s)}
\frac{\partial (N\Phi)}{\partial\theta_j(s^\prime)}\right]\right|_{\rm saddle}
\ed
\vspace*{-10mm}

\begin{equation}
~~~~~~~~~~~~~~~~~~~~
=-\delta_{ij}\avg{\hh(s)\hh(s^\prime)}_i
\label{fifth_link}
\end{equation} 
Note that we have already used identity (\ref{second_link}) to simplify
(\ref{fourth_link}) and (\ref{fifth_link}).

\section{Derivation of Saddle Point Equations}

In the limit $N\goesto\infty$, the integral (\ref{decomposed}) 
will be dominated by the dominating (physical) saddle point of the 
extensive exponent $\Psi\plus\Phi\plus\Omega$. 
We are now in a position to derive the saddle-point equations by
differentiation with respect to  our integration variables 
$\{\tm,\tk,\tmh,\tkh,\bq,\bQ,\bK,\bqh,\bQh,\bKh\}$. 
These equations will involve the average sequence overlap $m(s)$
(which measures the quality of the sequence recall) 
and the average single-site correlation- and response
functions $C(s,s^\prime)$ and $G(s,s^\prime)$: 
\begin{equation}
m(s)=\lim_{N\to\infty}\frac{1}{N}\sum_i \overline{\avg{\sigma_i(s)}}\xi_i^s
\label{overlap_def}
\end{equation}
\vspace*{-10mm}

\begin{equation}
C(s,s^\prime)=\lim_{N\to\infty}\frac{1}{N}\sum_i\overline{\avg{\sigma_i(s)\sigma_i(s^\prime)}}
\label{correlation_def}
\end{equation}
\vspace*{-10mm}

\begin{equation}
G(s,s^\prime)=\lim_{N\to\infty}\frac{1}{N}\sum_i\frac{\partial\overline{\avg{\sigma_i(s)}}}{\partial\theta_i(s^\prime)}
\label{response_def}
\end{equation}
Straighforward differentiation, followed by usage of the identities
(\ref{first_link}-\ref{fifth_link}) wherever possible, then
leads us to the following saddle-point equations:  
\begin{equation}
{\rm for~all~}s,s^\prime:~~~~~~
k(s)=\mh(s)=Q(s,s^\prime)=0
\label{saddle_1}
\end{equation}
\vspace*{-10mm}

\begin{equation}
{\rm for~all~}s:~~~~~~~~~~~
\kh(s)=m(s)=\lim_{N\to\infty}\frac{1}{N}\sum_i \avg{\sigma(s)}_i\xi_i^s
\label{saddle_2}
\end{equation}
\vspace*{-10mm}

\begin{equation}
{\rm for~all~}s,s^\prime:~~~~~~
q(s,s^\prime)=C(s,s^\prime)=
\lim_{N\to\infty}\frac{1}{N}\sum_i \avg{\sigma(s)\sigma(s^\prime)}_i
\label{saddle_3}
\end{equation}
\vspace*{-10mm}

\begin{equation}
{\rm for~all~}s,s^\prime:~~~~~~
K(s,s^\prime)=iG(s,s^\prime)=
\lim_{N\to\infty}\frac{1}{N}\sum_i \avg{\sigma(s)\hh(s^\prime)}_i
\label{saddle_4}
\end{equation}
\vspace*{-10mm}

\begin{equation}
{\rm for~all~}s,s^\prime:~~~~~~
\hat{q}(s,s^\prime)=\left.
 \frac{\rmi\partial\Omega}{\partial q(s,s^\prime)}\right|_{\rm
saddle}
\label{saddle_5}
\end{equation}
\vspace*{-10mm}

\begin{equation}
{\rm for~all~}s,s^\prime:~~~~~~
\hat{Q}(s,s^\prime)=\left.
 \frac{\rmi\partial\Omega}{\partial Q(s,s^\prime)}\right|_{\rm
saddle}
\label{saddle_6}
\end{equation}
\vspace*{-10mm}

\begin{equation}
{\rm for~all~}s,s^\prime:~~~~~~
\hat{K}(s,s^\prime)=\left.
 \frac{\rmi\partial\Omega}{\partial K(s,s^\prime)}\right|_{\rm saddle}
\label{saddle_7}
\end{equation}
The effective single-site measure (\ref{single_site_measure})
simplifies considerably due to (\ref{saddle_1},\ref{saddle_2}), and
since the function $\Omega$ 
depends on the trio
$\{\bq,\bQ,\bK\}$ only (see (\ref{define_omega})), 
our saddle-point equations can be  reduced to
a problem involving only the key physical observables
$m(s)$, $C(s,s^\prime)$ and $G(s,s^\prime)$. 

In order to calculate the remaining Gaussian integral in $\Omega$ we
have to define matrices operating in the product space of vectors
labelled by both time indices $s$ and pattern indices $\mu$. Note: 
in  the case of the standard symmetric Hopfield model, 
where integration variables with  different pattern labels $\mu$
immediately decouple, this would not have been
necessary.  
We define a matrix 
$\bGamma=\mS\otimes \mR$ as  having matrix elements
$\Gamma_{\mu\mu^\prime}(s,s^\prime)=S_{\mu\mu^\prime}R(s,s^\prime)$, 
where $\mu,\mu^\prime=1,\ldots,p$ and where
$s,s^\prime=0,\ldots,t\minus 1$.
It will operate as follows: if $\vy=\bGamma\vx$ then 
$y_\mu(s)=\sum_{\mu^\prime>t}\sum_{s^\prime<t}S_{\mu\mu^\prime}R(s,s^\prime)
x_{\mu^\prime}(s^\prime)$ for each $(\mu,s)$. 
Note that in evaluating $\Omega$ for $N\to\infty$, and due to $t$
remaining finite, we can 
safely drop the 
restriction that $\mu>t$, and instead have $\mu=1,\ldots,p$.   
The above definition allows
us to write
\bd
\lim_{N\to\infty}
\Omega[\bC,\bQ,\rmi\bG]=
\lim_{N\to\infty}\frac{1}{N}\ln\int\!
\frac{\cD{\tu}\cD{\tv}}{(2\pi)^{pt}}~
\rme^{~-\frac{1}{2}\vu\cdot[\mone\otimes\bQ]\vu
-\frac{1}{2}\vv\cdot[\mone\otimes\bC]\vv
+\rmi\vv\cdot[\bS\otimes\mone-\bG\otimes\mone]\vu}
\ed
\vspace*{-10mm}

\bd
=\lim_{N\to\infty}
\frac{1}{N}\ln\left\{{\rm det}^{-\frac{1}{2}}[\mone\otimes\bC]~
\int\!
\frac{\cD{\tu}}{(2\pi)^{pt/2}}~
\rme^{~-\frac{1}{2}\vu\cdot\left\{
\mone\otimes\bQ+
[\bS\otimes\mone-\mone\otimes\bG]^{\dag}
[\mone\otimes\bC]^{-1}
[\bS\otimes\mone-\mone\otimes\bG]\right\}\vu}\right\}
\ed
\vspace*{-10mm}

\begin{equation}
=
-\lim_{N\to\infty}\frac{1}{2N}\left\{\room
\ln~{\rm det}[\mone\!\otimes\!\bC]
+\ln~{\rm det}\left\{
\mone\!\otimes\!\bQ\plus 
[\bS\!\otimes\!\mone\minus \mone\!\otimes\!\bG]^{\dag}
[\mone\!\otimes\!\bC]^{-1}
[\bS\!\otimes\!\mone\minus \mone\!\otimes\!\bG]\right\}\right\}
\label{omega_integral}
\end{equation}
We use (\ref{omega_integral}) to work out the saddle-point
equations (\ref{saddle_5}-\ref{saddle_7}). The first of the three
equations comes out trivially:
\bd
\hat{q}(s,s^\prime)=\rmi
 \frac{\partial}{\partial C(s,s^\prime)}~\Omega[\bC,{\bf 0},\rmi\bG]
\ed
\vspace*{-10mm}

\begin{equation}
~~~~~~~~~~
=-\lim_{N\to\infty}\frac{\rmi}{2N}
 \frac{\partial}{\partial C(s,s^\prime)}~
\ln~{\rm det}\left\{ 
[\bS\!\otimes\!\mone\minus \mone\!\otimes\!\bG]^{\dag}
[\bS\!\otimes\!\mone\minus \mone\!\otimes\!\bG]\right\}
=0
\label{saddle_5_further}
\end{equation}
In order to work out the remaining two equations we use 
the general matrix identity $\ln {\rm det}[\bM\plus \bQ]=\ln {\rm
det}\bM+{\rm Tr}[\bM^{-1}\bQ]+{\cal O}(\bQ^2)$, as well as 
the specific properties of the $p\times p~$  
matrix $S_{\mu\nu}=\delta_{\mu,\nu+1}$. In particular we will be using 
its unitarity, 
 $\bS^\dag \bS=\mone$, the identity
$[(\bS^\dag)^m\bS^n]_{\mu\mu}=\delta_{mn}$, 
and its $p$ eigenvalues $s_\mu$ 
being given by  $s_\mu=e^{-2\pi\rmi \mu/p}$. Equation (\ref{saddle_6})
now reduces to
\bd
\hat{Q}(s,s^\prime)
=\rmi \lim_{\bQ\to{\bf 0}}
 \frac{\partial}{\partial Q(s,s^\prime)}~\Omega[\bC,\bQ,\rmi\bG] 
\ed
\vspace*{-10mm}

\bd
=
-\lim_{N\to\infty}\frac{\rmi}{2N}
\sum_{\mu\leq p}
\left\{
[\bS\!\otimes\!\mone\minus \mone\!\otimes\!\bG]^{\dag}
[\mone\!\otimes\!\bC]^{-1}
[\bS\!\otimes\!\mone\minus
\mone\!\otimes\!\bG]\right\}^{-1}_{\mu\mu}(s^\prime,s)
\ed
\vspace*{-10mm}

\bd
=
-\lim_{N\to\infty}\frac{\rmi}{2N}
\sum_{\mu\leq p}
\left\{
[\mone\!\otimes\!\mone\minus\bS^\dag\!\otimes\!\bG]^{-1}
[\mone\!\otimes\!\bC]
[\mone\!\otimes\!\mone\minus \bS\!\otimes\!\bG^\dag]^{-1}
\right\}_{\mu\mu}(s^\prime,s)
\ed
\vspace*{-10mm}

\bd
=
-\frac{1}{2}\alpha \rmi\sum_{n,m\geq 0}
\lim_{p\to\infty}\frac{1}{p}
\sum_{\mu\leq p}
\left\{
[\bS^\dag\!\otimes\!\bG]^n
[\mone\!\otimes\!\bC]
[\bS\!\otimes\!\bG^\dag]^m
\right\}_{\mu\mu}(s^\prime,s)
\ed
giving:
\begin{equation}
\hat{\bQ}=
-\frac{1}{2}\alpha \rmi\sum_{n\geq 0}
(\bG^\dag)^n \bC (\bG)^n
\label{saddle_6_further}
\end{equation}
Finally we turn to equation (\ref{saddle_7}):
\bd
\hat{K}(s,s^\prime)
= \frac{\partial}{\partial G(s,s^\prime)}~\Omega[\bC,{\bf 0},\rmi\bG]
\ed
\vspace*{-10mm}

\bd
=-\frac{1}{2}\alpha~
 \frac{\partial}{\partial G(s,s^\prime)}~
\lim_{p\to\infty}\frac{1}{p}\left\{
\ln~{\rm det}
[\mone\!\otimes\!\mone\minus \bS^{\dag}\!\otimes\!\bG]^{\dag}
+\ln~{\rm det}
[\mone\!\otimes\!\mone\minus \bS^{\dag}\!\otimes\!\bG]
\right\}
\ed
\vspace*{-10mm}

\bd
=-\frac{1}{2}\alpha~
 \frac{\partial}{\partial G(s,s^\prime)}~
\lim_{p\to\infty}\frac{1}{p}\sum_{\mu\leq p}
\left\{
\ln~{\rm det}
[\mone\minus \rme^{-2\pi\rmi\mu/p} \bG^{\dag}]
+\ln~{\rm det}
[\mone\minus \rme^{2\pi\rmi\mu/p} \bG]
\right\}
\ed
\vspace*{-10mm}

\begin{equation}
=-\frac{1}{2}\alpha~
 \frac{\partial}{\partial G(s,s^\prime)}~
{\rm Tr}\sum_{n>0}
\frac{1}{n}
\int_{-\pi}^{\pi}\!\frac{\rmd\omega}{2\pi}
\left\{
 \rme^{-n\rmi\omega} (\bG^{\dag})^n 
+
\rme^{n\rmi\omega} (\bG)^n
\right\}=0
\label{saddle_7_further}
\end{equation}

\section{The Effective Single-Spin Problem}

Let us summarize the present stage of our calculation. Most
macroscopic integration variables are found to vanish in the relevant
physical 
saddle-point:
$k(s)\!=\!\hat{m}(s)\!=\!Q(s,s^\prime)\!=\!\hat{q}(s,s^\prime)\!=\!
\hat{K}(s,s^\prime)\!=\!0$.
The remaining ones can all be expressed in terms of
three macroscopic observables, viz. the
overlaps $m(s)$ and the single-site correlation and response functions
$C(s,s^\prime)$ and $G(s,s^\prime)$, as defined in
(\ref{overlap_def},\ref{correlation_def},\ref{response_def}), 
by using the four 
equations
(\ref{saddle_2},\ref{saddle_3},\ref{saddle_4},\ref{saddle_6_further}).
We are thus left with a set of closed equations (\ref{saddle_2},\ref{saddle_3},\ref{saddle_4}) from which to solve  
$\{m(s),C(s,s^\prime),G(s,s^\prime)\}$. These equations 
are defined in terms of
an effective 
single-spin problem. 
At this stage it is natural to choose the
remaining external fields $\theta_i(s)$ to be so-called `staggered'
ones, i.e. $\theta_i(s)=\theta(s)\xi_i^{s+1}$. If used as
symmetry-breaking perturbations, such fields will exactly 
single out macroscopic solutions of the type we introduced as an ansatz. 
This choice also removes the formal need to break symmetries via initial
conditions, so that we may now choose $p_i(\sigma(0))=p(\sigma(0))$.  
As a consequence we find that the single-site measure 
(\ref{single_site_measure}) 
becomes site-independent, since the
remaining site dependence due to pattern components $\xi_i^\mu$ can be
eliminated via a gauge transformation  
whereby $\sigma(s)\to
\sigma(s)\xi_i^s$ and $h(s)\to h(s)\xi_i^{s+1}$. 
The resulting single-spin problem involves the following measure
(which is properly normalized, as can be verified by explicit
evaluation of $\avg{1}_{\star}$): 
\bd
\avg{f[\{\sigma\}]}_{\star}= 
\sum_{\sigma(0)\ldots\sigma(t)}
\int\!\{\rmd\rmh \rmd\rmhh\}~ p(\sigma(0))~f[\{\sigma\}]~
\rme^{~\sum_{s<t}\left[\beta\sigma(s+1)h(s)-\ln 2\cosh[\beta h(s)]\right]}
\ed
\vspace*{-11mm}

\begin{equation}
~~~~~~~~~~~~~~~~~~~~\times~
\rme^{\rmi\sum_{s<t}\hh(s)[h(s)-\theta(s)-m(s)]
-\frac{1}{2}\alpha\sum_{s,\stick<t}R(s,\stick)\hh(s)\hh(\stick)}
\label{single_spin}
\end{equation}
with $R(s,s^\prime)=\sum_{n\geq 0}[(\bG^\dag)^n\bC(\bG)^n](s,s^\prime)$. 
This measure describes an effective single spin $\sigma(s)$ with a 
stochastic alignment to 
local fields given by $h(s)=m(s)\plus \theta(s)\plus \alpha^{\frac{1}{2}}\phi(s)$, in
which the term $\phi(s)$ represents a zero-average Gaussian random
field with (non-zero) temporal correlations
$\avg{\phi(s)\phi(s^\prime)}= R(s,s^\prime)$. 
Note that, as a consequence of (\ref{saddle_7_further}), 
there is no term representing a retarded
self-interaction, in contrast with the standard (symmetric) Hopfield
model. 
This is the mathematical explanation of the 
higher storage capacity in the present sequence processing model. 
The asymmetry of the  interaction matrix
prevents the build-up of a microscopic memory, similar to 
the situation in the non-symmetric 
SK--model \cite{crisanti87,crisanti88,rieger89}.  
The equations from which to solve our remaining order parameters
can be written as
\begin{equation}
m(s)=\avg{\sigma(s)}_{\star}
\label{m_eff}
\end{equation} 
\vspace*{-10mm}

\begin{equation}
C(s,s^\prime)=\avg{\sigma(s)\sigma(s^\prime)}_{\star}
\label{C_eff}
\end{equation} 
\vspace*{-10mm}

\begin{equation}
G(s,s^\prime)=\frac{\partial}{\partial\theta(s^\prime)}\avg{\sigma(s)}_{\star}
\label{G_eff}
\end{equation} 
Since the measure (\ref{single_spin}) factorizes with respect to spin
variables at different times, we can immediately perform the spin
summations
in (\ref{m_eff},\ref{C_eff},\ref{G_eff}) (which would not
have been possible for the standard Hopfield model). After a simple
rescaling of fields and conjugate fields we then arrive at
\begin{equation}
m(s)=
\int\!\{\rmd\tv\rmd\tw\}~\rme^{\rmi\tv\mal\tw-\half\tw\mal\mR\tw}
~\tanh \beta[m(s\minus 1)\plus \theta(s\minus
1)\plus\alpha^{\frac{1}{2}} v(s\minus 1)]
\label{m_macro}
\end{equation}
\bd
C(s,s^\prime)=\delta_{s,s^\prime}+[1\minus \delta_{s,s^\prime}]
\int\!\{\rmd\tv\rmd\tw\}~\rme^{\rmi\tv\mal\tw-\half\tw\mal\mR\tw}
~\tanh \beta[m(s\minus 1)\plus \theta(s\minus
1)\plus\alpha^{\frac{1}{2}} v(s\minus 1)]
\ed
\vspace*{-11mm}

\begin{equation}
~~~~~~~~~~~~~~~~~~~~
\times ~
\tanh \beta[m(s^\prime\minus 1)\plus \theta(s^\prime\minus
1)\plus \alpha^{\frac{1}{2}} v(s^\prime\minus 1)]
\label{C_macro}
\end{equation}
\vspace*{-11mm}

\begin{equation}
G(s,s^\prime)=\beta\delta_{s,s^\prime+1}\left\{
1-\!\int\!\{\rmd\tv\rmd\tw\}~\rme^{\rmi\tv\mal\tw-\half\tw\mal\mR\tw}
~\tanh^2 \beta[m(s\minus 1)\plus \theta(s\minus
1)\plus \alpha^{\frac{1}{2}} v(s\minus 1)]
\right\}
\label{G_macro}
\end{equation}
with $R(s,s^\prime)=\sum_{n\geq 0}[(\bG^\dag)^n\bC(\bG)^n](s,s^\prime)$. 
The response function is found to be non-zero only if field perturbation and
spin measurement are temporally separated by exactly one iteration step. Thus
anomalous response cannot occur, and macroscopic stationarity should
be achieved on finite time-scales.

\section{The Stationary State}

We now choose stationary external fields $\theta_i(s)=\theta
\xi_i^{s+1}$, giving $\theta(s)=\theta$ in terms of the single-spin problem, 
and inspect time-translation invariant solutions of our macroscopic 
equations (\ref{m_macro},\ref{C_macro},\ref{G_macro}), which will describe
motion on a macroscopic limit cycle:
\begin{equation}
m(s)=m~~~~~~~~~~
C(s,\stick)=C(s-\stick)
~~~~~~~~~~
G(s,\stick)=G(s-\stick)
\label{stationary}
\end{equation}
In order to do this we shift the initial time in
(\ref{m_macro},\ref{C_macro},\ref{G_macro}) from $t_0=0$ to
$t_0=-\infty$, and the final time to $t=\infty$.  
According to (\ref{stationary}) the matrices $\mC$ and $\mG$ 
become Toeplitz matrices and
commute, which implies that the matrix $\mR$ simplifies to
\begin{equation}
\mR=\mC\left[\mone-\mG^\dagger\mG\right]^{-1},~~~~~~~~~~
R(s,\stick)=R(s\minus \stick)
\label{stationary_R}
\end{equation}
and that we may thus write the stationary 
version of (\ref{m_macro},\ref{C_macro},\ref{G_macro})
as 
\bd
m=
\int\!\{\rmd\tv\rmd\tw\}~\rme^{\rmi\tv\mal\tw-\half\tw\mal\mR\tw}
~\tanh \beta[m\plus \theta \plus\alpha^{\frac{1}{2}} v(0)]
\ed
\vspace*{-11mm}

\bd
C(\tau\!\neq\! 0)=
\int\!\{\rmd\tv\rmd\tw\}~\rme^{\rmi\tv\mal\tw
-\half\tw\mal\mR\tw}
~\tanh \beta[m\plus \theta\plus\alpha^{\frac{1}{2}} v(\tau)]
\tanh \beta[m\plus \theta\plus \alpha^{\frac{1}{2}} v(0)]
\ed
\vspace*{-11mm}

\bd
G(\tau)=\beta\delta_{\tau,1}\left\{
1-\!\int\!\{\rmd\tv\rmd\tw\}~\rme^{\rmi\tv\mal\tw-\half\tw\mal\mR\tw}
~\tanh^2 \beta[m\plus \theta\plus \alpha^{\frac{1}{2}} v(0)]
\right\}
\ed
We separate in $C(\tau)$ and $R(\tau)$ the persistent from the
non-persistent parts, i.e.
\bd
C(\tau)=q+\tilde{C}(\tau),~~~~~~~~~~
R(\tau)=r+\tilde{R}(\tau),~~~~~~~~~~
\lim_{\tau\to\pm\infty}\tilde{C}(\tau)=\lim_{\tau\to\pm\infty}\tilde{R}(\tau)=0
\ed
The persistent part $r$ of $R(\tau)$ can be expressed in terms of the
persistent part $q$ of $C(\tau)$, by combining (\ref{stationary_R}) with
the above expression for $G(\tau)$.  
This separation of persistent parts induces 
a frozen random field into the above order parameter
equations, which can subsequently be absorbed into the local fields:
\bd
\rme^{\rmi\tv\mal\tw-\half\tw\mal\mR\tw}=
\rme^{\rmi\tv\mal\tw-\half r\left[\sum_{s}w(s)\right]^2-\half\tw\mal\tilde{R}\tw}
=
\int\!Dz~\rme^{\rmi\sum_s w(s)[v(s)- z\sqrt{r}]-\half\tw\mal\tilde{R}\tw}
\ed
(with the familiar abbreviation
$Dz=(2\pi)^{-\frac{1}{2}}\rme^{-\frac{1}{2}z^2}$). 
Upon rewriting $G(\tau)=\beta\delta_{\tau,1}\left[1-\tilde{q}\right]$
and $r=q\rho$, we
arrive at the following expressions for our persistent observables: 
\bd
m=
\int\!Dz \int\!\{\rmd\tv\rmd\tw\}~\rme^{\rmi\tv\mal\tw-\half\tw\mal\tilde{\mR}\tw}
~\tanh \beta[m\plus \theta \plus z\sqrt{\alpha q\rho }
\plus\alpha^{\frac{1}{2}} v(0)]
\ed
\vspace*{-11mm}

\bd
q=\lim_{\tau\to\infty}\int\!Dz
\int\!\{\rmd\tv\rmd\tw\}~\rme^{\rmi\tv\mal\tw-\half\tw\mal\tilde{\mR}\tw}
~\tanh \beta[m\plus \theta\plus z\sqrt{\alpha
q\rho}\plus\alpha^{\frac{1}{2}} v(\tau)]
\ed
\vspace*{-13mm}

\bd
~~~~~~~~~~~~~~~~~~~~~~~~~~~~~~~~~~~~~~~~
\times~
\tanh \beta[m\plus \theta\plus z\sqrt{\alpha q\rho}\plus \alpha^{\frac{1}{2}} v(0)]
\ed
\vspace*{-11mm}

\bd
\tilde{q}=\int\!Dz
\int\!\{\rmd\tv\rmd\tw\}~\rme^{\rmi\tv\mal\tw-\half\tw\mal\tilde{\mR}\tw}
~\tanh^2 \beta[m\plus \theta\plus z\sqrt{\alpha q\rho}
\plus \alpha^{\frac{1}{2}} v(0)]
\ed
\vspace*{-11mm}

\bd
\rho=\left[1\minus \beta^2(1-\tilde{q})^2\right]^{-1}
\ed
We only need to know the joint probability distribution of the
pair $(v(\tau),v(0))$ in the limit $\tau\goesto\infty$ to work out the
remaining integrals over $\tv$ and $\tw$. This distribution is clearly
a  zero-average Gaussian one, so finding the second order moments
suffices.  Integration over $\tw$ gives
\[
\avg{v(\tau)v(0)}=[{\rm det}\tilde{\mR}]^{-\frac{1}{2}}
\int\!\prod_s\left[\frac{\rmd v(s)}{\sqrt{2\pi}}\right]~
\rme^{-\half \tv\mal\mRt^{-1}\tv}~v(\tau)v(0)=\Rt(\tau)
\]
from which 
we conclude that $\avg{v(0)^2}=\Rt(0)$, and that 
$\lim_{\tau\goesto\infty}\avg{v(\tau)v(0)}=0$. The variance 
$\Rt(0)=R(0)\minus r$ immediately follows from (\ref{stationary_R}):
\[
\Rt(0)=\frac{1-q}{1-\beta^2(1\minus \tilde{q})^2}=(1\minus q)\rho
\]
All remaining integrals are now expressed in terms of 
persistent observables only:
\bd
m=
\int\!Dz \int\!Dx
~\tanh \beta[m\plus \theta \plus z\sqrt{\alpha q\rho}\plus
x\sqrt{\alpha(1\minus q)\rho}]
\ed
\vspace*{-11mm}

\bd
q=
\int\!Dz \left[\int\! Dx
~\tanh \beta[m\plus \theta \plus z\sqrt{\alpha q\rho}\plus
x\sqrt{\alpha(1\minus q)\rho}]\right]^2
\ed
\vspace*{-11mm}

\bd
\tilde{q}=
\int\!Dz \int\! Dx
~\tanh^2 \beta[m\plus \theta \plus z\sqrt{\alpha q\rho}\plus
x\sqrt{\alpha(1\minus q)\rho}]
\ed
If we finally 
combine the two Gaussian variables in the equations for $m$ and $\tilde{q}$ 
 into a single Gaussian variable we
arrive at our final result:
\begin{equation}
\rho=\left[1\minus \beta^2(1-\tilde{q})^2\right]^{-1}
\label{rho_final}
\end{equation}
\vspace*{-11mm}

\begin{equation}
m=
\int\!Dz 
~\tanh \beta[m\plus \theta \plus z\sqrt{\alpha \rho}]
\label{m_final}
\end{equation}
\vspace*{-11mm}

\begin{equation}
\tilde{q}=
\int\!Dz 
~\tanh^2 \beta[m\plus \theta \plus z\sqrt{\alpha \rho}]
\label{qtilde_final}
\end{equation}
\vspace*{-11mm}

\begin{equation}
q=
\int\!Dz \left[\int\! Dx
~\tanh \beta[m\plus \theta \plus z\sqrt{\alpha q\rho}\plus
x\sqrt{\alpha(1\minus q)\rho}]\right]^2
\label{q_final}
\end{equation}
Note that the trio (\ref{rho_final},\ref{m_final},\ref{qtilde_final})
form itself a closed set, from the solution of which the persistent
correlation $q$ simply follows.

\section{Phase Diagram and Storage Capacity}

\begin{figure}[t]
\vspace*{85mm}
\hbox to\hsize{\hspace*{25mm}\includegraphics{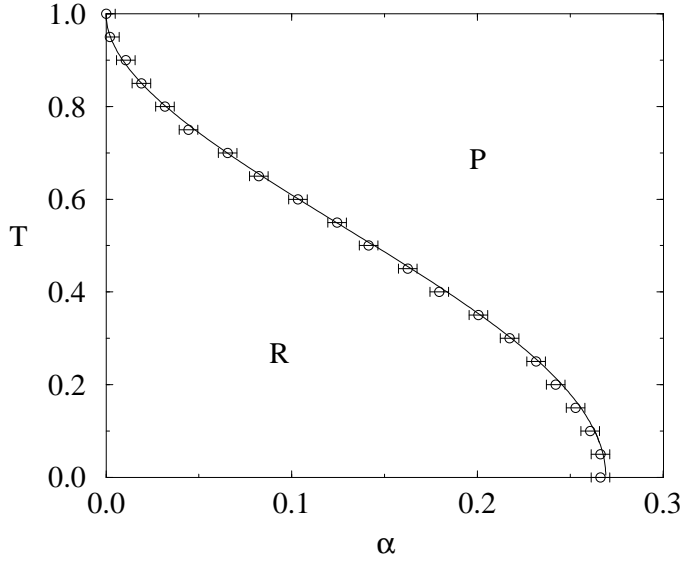}\hspace*{-25mm}}
\vspace*{-4mm}
\caption{Phase diagram of the sequence processing network, in which one
finds two phases: a recall phase (R), 
characterized by $\{m\neq 0,~q>0,~\tilde{q}>0\}$, and a
paramagnetic phase (P), characterized by $\{m=0,~q=0,~\tilde{q}>0\}$. 
Solid line: the theoretical
prediction for the phase boundary. 
Markers: simulation results for systems of $N=10,000$ spins
measured after $2,500$ iteration steps. The precision in terms of $\alpha$
is at least $\Delta\alpha=0.005$ (indicated by error bars); the values
for $T$ are exact.}
\label{finaldiag}
\end{figure}

We have solved the coupled 
equations (\ref{rho_final}--\ref{qtilde_final}) numerically for
$\theta=0$ 
\footnote{The alternative choice  $\theta\neq 0$ 
would have described the 
less interesting scenario where the $m\neq 0$ state 
would not be sustained autonomously (if at
all), 
but where 
at each time step and at each site a very specific external field
$\theta_i(s)=\theta\xi_i^{s+1}$ would have actively pushed the
system towards the pattern sequence.}   
in order to 
determine the region in the $\alpha$--$T$ plane where solutions with
$m\neq 0$, which describe pattern sequence recall, exists. 
The boundary of 
this region  determines the storage capacity of the system. This
theoretical result was tested against numerical 
 simulations of the present model, carried out at the spin level
(\ref{flip_prob}). We show  the results in figure
\ref{finaldiag}.  One finds that, for $T>0$ and $\alpha<\infty$, the
equations (\ref{rho_final}--\ref{qtilde_final}) admit only two types of
solutions: a recall solution (R) characterized by $\{m\neq
0,~q>0,~\tilde{q}>0\}$, and a paramagnetic solution (P) characterized
by $\{m=0,~q=0,~\tilde{q}>0\}$. The absence of the analogon of a
spin-glass phase will be discussed in more detail below.  The phase
boundary R$\to$P as obtained theoretically (solid line) shows an
excellent agreement with the computer simulations (markers), as
performed for systems of size $N=10,000$ (using bi-section). 
The maximum storage capacity $\alpha_\crit$  is obtained in the zero
noise limit $T\to 0$ (or $\beta\to\infty$). For $\beta\to\infty$, where
$\tilde{q}\to 1$ and $q\to 1$, the
saddle point equations can be simplified in the usual manner, using
identities such as
\bd
\lim_{\beta\goesto\infty}
\int\rmD z~\tanh\beta\left[m+z\sqrt{\alpha\rho}\right]=
\erf\left[\frac{m}{\sqrt{2\alpha\rho}}\right]
\ed
\vspace*{-10mm}

\begin{figure}[t]
\vspace*{85mm}
\hbox to\hsize{\hspace*{25mm}\includegraphics{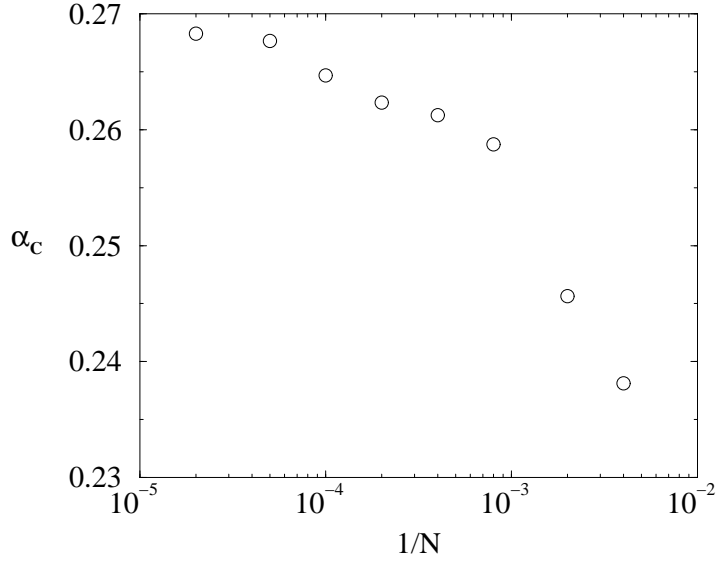}\hspace*{-25mm}}
\vspace*{-4mm}
\caption{Results of determining the maximum sequence storage capacity
$\alpha_\crit$ at $T=0$ via numerical simulation of networks with
different sizes $N$.  The values for $\alpha_\crit$ have been
determined with a precision of at least
$\Delta\alpha=0.001$ where possible.  Note that the $N=\infty$ theory
predicts $\alpha_\crit\approx 0.269$.}
\label{scaling} 
\end{figure}

\bd
\lim_{\beta\goesto\infty}
\beta(1-\qt)=\frac{\partial}{\partial m}\lim_{\beta\goesto\infty}
\int\rmD z~\tanh\beta\left[m+z\sqrt{\alpha\rho}\right]
=\sqrt{\frac{2}{\pi\alpha\rho}}~\exp\left[-\frac{m^2}{2\alpha\rho}\right]
\ed
With the definition $x=m/\sqrt{2\alpha\rho}$, from which the overlap 
$m$ follows according to $m={\rm erf}(x)$, 
we can combine our saddle point equations for $\beta\to\infty$ into
the single transcendental equation
\begin{equation}
x\sqrt{2\alpha}=\pm\sqrt{{\rm erf}^2(x)-\frac{4x^2}{\pi}\exp\left(-2x^2\right)}.
\label{eq:transcendental}
\end{equation}
This equation is identical to that obtained in the  $T=0$ limit for
the   layered model of \cite{domany89},  and for the present model we thus 
obtain the same maximum storage 
capacity, which is defined as the largest value of $\alpha$ for which
(\ref{eq:transcendental}) has non-trivial solutions, 
of $\alpha_\crit\approx 0.269$. Note, however, that this
equivalence does not extend beyond the $T=0$ limit. 
To also verify this latter result with numerical simulations, 
taking into account the possibility of finite size effects, 
we measured the maximum storage
capacity in zero temperature simulations for different system sizes, ranging from $N=2,500$ to
$N=50,000$. This resulted in figure \ref{scaling}. 
The numerical data are again perfectly consistent with the result
$\alpha_\crit\approx 0.269$ of
our $N=\infty$ theory.  

Finally we turn to the non-recall phases, still for zero external field, 
where $m=0$ and where the
remaining order parameters $q\in[0,1]$ and $\tilde{q}\in[0,1]$ follow from solving
the coupled equations
\begin{eqnarray}
\qt&=\int\rmD z~\tanh^2\left[\beta z\sqrt{
        \frac{\alpha}{1-\beta^2(1-\qt)^2}}\right]
\label{eq:para1}
\\
q&=\int\rmD z\left\{ \int\rmD x ~\tanh\beta\left[
        z\sqrt{\frac{\alpha q}{1-\beta^2(1-\qt)^2}}+
        x\sqrt{\frac{\alpha(1-q)}{1-\beta^2(1-\qt)^2}}\right]
        \right\}^2
\label{eq:para2}
\end{eqnarray}
The first of these equations (\ref{eq:para1}) 
determines $\tilde{q}$, which is related to the response function via 
$G(\tau)=\beta(1-\tilde{q})\delta_{\tau,1}$. Its solution is unique.   
For finite temperature one finds that $\tilde{q}$ is always non-zero, 
approaching zero only asymptotically as 
$\tilde{q}=\alpha\beta^2+\Order{\beta^4}$ 
for $T\to\infty$. The persistent
correlation $q$ subsequently follows from solving (\ref{eq:para2}). 
This second equation always admits the paramagnetic solution $q=0$. Careful
numerical and analytical inspection reveals that for $T>0$ and
$\alpha<\infty$ it admits no solutions with $q>0$, which would have
been the analogon of a spin-glass state. 
Only in the limits $T\to 0$ and $\alpha\to\infty$, 
where $\beta(1-\tilde{q})\to (1+\frac{1}{2}\pi\alpha)^{-\frac{1}{2}}$
and $\beta(1-\tilde{q})\to 1$ respectively, and where equation   
(\ref{eq:para2}) converts into 
\[
q=\int\rmD z~{\rm erf}^2\left[\frac{z\sqrt{q}}{\sqrt{2(1-q)}}\right]
\]
does one find a non-trivial solution, namely $q=1$. 
This implies that in the phase diagram of figure \ref{finaldiag} the
phase beyond the boundary of the recall region is a paramagnetic
state, with only a transition to a spin-glass type frozen state
precisely  at $T=0$. This type of behaviour is very similar to that
observed in non-symmetric spin-glass models
\cite{crisanti87,crisanti88,rieger89}. 

\section{Discussion}

In this paper we have used path integral methods 
to solve in the thermodynamic limit the
dynamics of a non-symmetric neural network model, designed to store
and recall sequences of stored patterns, close to saturation.  For
about a decade this model has been known from numerical simulations to
have a significantly enlarged storage capacity (by about a factor two)
compared to the more familiar symmetric Hopfield network
\cite{hopfield82,amit85,amit87}, which stores static patterns and obeys
detailed balance.  So far the sequence processing 
model had not yet been solved, and thus 
the enlarged storage capacity had not yet been explained, mainly due
to the complication that the absence of detailed balance
rules out the more traditional equilibrium statistical mechanical
methods, including replica theory.  In contrast, even in the regime of
interest where the number of patterns in the sequence scales as
$p=\alpha N$, and where thus the dynamical methods of simple mean-field
models cannot be used, the powerful path integral methods of
\cite{dedominicis78,sommers87,rieger88} do still apply; they allow us
perform the disorder average in a dynamical framework, and thereby to
calculate the system's phase diagram without having to resort to
additional approximations. 

In the standard
(symmetric) Hopfield network two effects limit the storage capacity: a
Gaussian noise in the equivalent effective single spin problem, which
is non-local in time, and a retarded self-interaction. The magnitude
of both contributions depends on the load factor $\alpha$. For the
present model  we find, in
contrast, 
that the retarded self-interaction vanishes, 
which explains the extended storage capacity. Numerical
simulations for large system sizes (up to $N=50,000$ spins) are in
excellent agreement with our analytical results, both with respect to
the maximum storage capacity $\alpha_{\text{c}}\approx 0.269$ (at zero 
noise level) and with 
respect to the full phase
diagram in the $\alpha-T$ plane. 
In the limit of zero noise level we find that the equation from which to solve the 
order parameter which describes the quality of the sequence recall 
reduces to that of the layered model of \cite{domany89}.  
Our order parameter equations and their solutions also turn out to be 
very similar to those found for  various versions of the non-symmetric  
SK spin-glass model, as studied in 
\cite{crisanti87,crisanti88,rieger89}. In particular, common
features are the absence of a retarded self-interaction in the
effective single-spin problem, and  the absence of a spin-glass type 
phase for non-zero temperatures. 

As a next step one could apply the present formalism to networks
which store more than one periodic pattern sequence. By varying the
scaling with $N$ of 
both the sequence length and of the number of sequences, one should expect a
transition between the behaviour similar to that of the symmetric
Hopfield model (with an effective retarded self-interaction) and the
behaviour observed in the present model (without such a retarded
self-interaction). This will be the subject of a future study.

\section*{References}

\bibliography{abbrev,spinsys,general-physics,hopfield,neural-networks,unfollowed}

\end{document}